# Ni Schottky barrier on heavily doped phosphorous implanted 4H-SiC

**M. Vivona, G. Greco, M. Spera, P. Fiorenza, F. Giannazzo, A. La Magna and F. Roccaforte**

*Consiglio Nazionale delle Ricerche – Istituto per la Microelettronica e Microsistemi (CNR_IMM), Strada VIII, n. 5 Zona Industriale, 95121 Catania, Italy*

**Abstract**

The electrical behavior of Ni Schottky barrier formed onto heavily doped ($N_D > 10^{19}$ cm$^{-3}$) n-type phosphorous implanted silicon carbide (4H-SiC) was investigated, with a focus on the current transport mechanisms in both forward and reverse bias. The forward current-voltage characterization of Schottky diodes showed that the predominant current transport is a thermionic-field emission mechanism. On the other hand, the reverse bias characteristics could not be described by a unique mechanism. In fact, under moderate reverse bias, implantation-induced damage is responsible for the temperature increase of the leakage current, while a pure field emission mechanism is approached with bias increasing. The potential application of metal/4H-SiC contacts on heavily doped layers in real devices are discussed.

## 1. Introduction

In last decades, wide-bandgap semiconductors have attracted great interest for the development of high-power electronic devices and systems [1,2]. The interest towards these materials is based on their extraordinary physical and electronic properties, such as a wide bandgap, high breakdown electric field strength, high saturation electron velocity and high thermal conductivity [3]. Particularly, the hexagonal polytype of silicon carbide (4H-SiC) plays a pivotal role thanks to the maturity reached in terms of material quality and technological implementation of the device fabrication steps [4]. In fact, 4H-SiC unipolar devices, such as Schottky barrier diodes (SBDs) and metal-oxide-semiconductor field effect transistors (MOSFETs), with high performances (low on-state voltage drop, high breakdown voltage, high switching speed, possibility to operate at high temperature, etc.), have become largely available on the market [4,5,6].

However, in order to fully exploit the potentialities of this material, significant efforts are currently devoted to address some physical concerns, which still limit the performances of SiC devices [7]. In this context, a full understanding of the behaviour of Schottky barriers on 4H-SiC is particularly useful in specific applications. For example, while in SBDs used as sensors and detectors a high Schottky barrier height is desired [8,9], in power electronics applications a strong attention is paid to the reduction of power consumption of

SBDs, through the exploration of different routes for a reduction of the barrier [10]. Typically, the metal/semiconductor barrier properties can be tailored by an optimized choice of metal system (e.g., low work-function metals, tunable compositions, etc.) [11,12,13,14], by employing suitable semiconductor surface treatments [15,16] or by intentionally changing the electric field distribution below the interface. As an example, ion-irradiation-induced damage below the interface in Ti/4H-SiC Schottky diodes showed the possibility to increase the barrier height by a de-activation of the dopant and a reduction of the electric field at the interface following a re-ordering of the crystal structure [17,18].

Ion-implantation is an essential process for both n-type and p-type selective doping in 4H-SiC devices [19]. In particular, heavily-doped n-type and p-type implanted regions are used in junction barrier Schottky (JBS) rectifiers and metal oxide semiconductor field effect transistors (MOSFETs) for the creation of Ohmic contacts [20,21,22]. On such implanted layers, Ohmic contacts can be formed by annealing at high temperature (> 900°C) of Nickel films, owing to the formation of nickel silicide ($Ni_2Si$) [22,23]. Hence, the knowledge of the carrier transport mechanisms at metal/heavily-doped SiC interfaces can be important for the optimization of the contact regions in these devices. Recently, Hara et al. [24] studied the forward carrier transport mechanism in Schottky diodes fabricated on 4H-SiC epitaxial layers with different donor concentrations (up to a doping concentration as high as

$1.8\times10^{19}$ cm$^{-3}$), observing a reduction of the barrier height with increasing the epilayer doping.

In this work, the electrical behavior of Ni Schottky contacts onto heavily doped ($N_D > 10^{19}$ cm$^{-3}$) n-type phosphorus-implanted 4H-SiC was investigated, in both forward and reverse bias. Specifically, a reduced turn-on voltage was achieved under forward bias, where the thermionic field emission mechanism dominates the current transport through the interface. On the other hand, the behavior of the contact under reverse bias was explained considering a tunneling mechanism assisted by the presence of ion-implantation-induced defects for moderate reverse bias, while probably approached a pure field emission regime with bias increasing. Finally, a numerical simulation of the potential distribution in a JBS diode structure demonstrated the feasibility of this approach in real devices.

## 2. Experimental details

The material used in our study was a n-type 4H-SiC epitaxial layer (with nitrogen doping of $1\times10^{16}$ cm$^{-3}$) grown onto a heavily doped 4H-SiC (0001) substrate. First, the upper part of the epitaxial layer was implanted at 400 °C using phosphorus (P) ions at energies ranging from 30 to 200 keV and with ion doses between $7.5\times10^{13} - 5\times10^{14}$ cm$^{-2}$. In this way, an implantation profile extending over 200 nm and with a peak concentration of $1\times10^{20}$ cm$^3$ is obtained, as confirmed by secondary ion mass spectrometry (SIMS) [25]. Then, a post-implantation thermal annealing treatment was carried out at 1675 °C in Ar ambient, protecting the surface with a carbon capping layer [26], in order to achieve the electrical activation of the dopant.

Schottky barrier diodes were fabricated on this material. Before the front-side processing of the sample, a large-area back-side contact was fabricated by Ni-deposition followed by rapid thermal annealing (RTA) at 950 °C in N$_2$-atmosphere [22]. Successively, a 100 nm-thick Ni-layer was deposited on the P-implanted surface by direct current (DC) magnetron sputtering and defined in circular structures (diameter of 250 μm) by optical lithography and lift-off process. These Ni-Schottky contacts were deliberately not subjected to annealing treatments to avoid interface reactions, which would induce a consumption of the SiC layer and/or a degradation of the barrier properties [27].

The Schottky contacts were characterized by means of current-voltage (I-V) measurements, under both forward and reverse bias, at various temperatures in the range 25-115 °C (step of 15 °C). The measurements were performed in a Karl-Suss MicroTec probe station equipped with a parameter analyzer.

## 3. Results and Discussion

Firstly, Fig.1 compares the forward current density-voltage (J-V$_F$) characteristics of Ni Schottky contacts formed on an n-type 4H-SiC epilayer, with or without a heavily-doped n-type implanted layer.

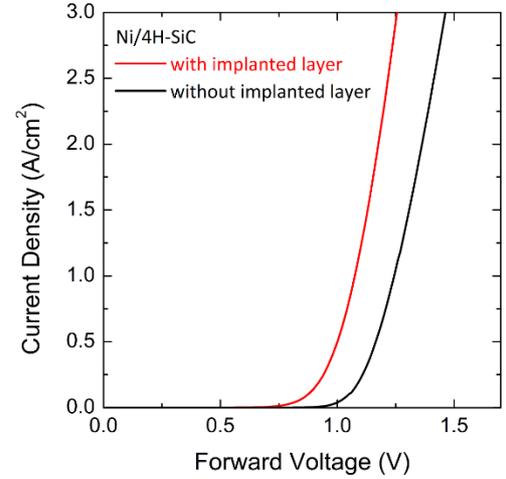

**Figure 1.** Forward J-V$_F$ characteristics of Ni Schottky contacts on an n-type 4H-SiC epilayer, with or without a heavily-doped n-type implanted layer.

As one can see, the J-V$_F$ curve of the contact on the heavily-doped n-type implanted 4H-SiC exhibts a lower turn-on voltage with respect to the reference contact formed on the 4H-SiC epilayer without implant.

Under forward bias, the electrical behavior of 4H-SiC Schottky diodes are typically described by the thermionic emission (TE) model [28,29], with the current density-voltage (J$_{TE}$ -V$_F$) relationship expressed by:

$$J_{TE} = A^*T^2 \times exp\left(-\frac{q\phi_{BTE}}{k_BT}\right) \times exp\left(q\frac{V_F-J_{TE}R_{on}}{nk_BT}\right) \ (1)$$

where $A^*$ is the effective Richardson's constant of 4H-SiC (146 A·cm$^{-2}$·K$^{-2}$) [30], $k_B$ = 1.38× $10^{-23}$ J/K is the Boltzmann's constant, q is the elementary charge, V$_F$ is the voltage applied across the metal/semiconductor interface and T is the absolute temperature. The relevant diode parameters in Eq. (1), i.e., ideality factor n, Schottky barrier height $\Phi_{BTE}$ and specific on-resistance $R_{ON}$, were derived as best fit parameters. The experimental value of barrier height measured in Ni-Schottky contacts on 4H-SiC epitaxial layer (with a doping concentration in the low $10^{16}$ cm$^{-3}$ range) typically ranges between 1.3 - 1.6 eV [10,31,32].

In our case, Ni Schottky contacts fabricated directly on the 4H-SiC epilayer (without implanted layer, black curve in Fig.1) resulted in a barrier height of 1.32eV,determined by using the TE model.

On the other hand, the TE model applied to the Ni Schottky contacts on the heavily-doped n-type implated layer gave a barrier height $\Phi_{BTE}$ = 0.94 eV and an ideality factor n = 1.8. Such a strong discrepancy with respect to the ideal behavior (n=1) suggests that the current transport at the Ni/heavily-doped 4H-SiC interface cannot be described by a pure TE regime. More reasonably, considering the high doping concentration of the implanted region, the presence of a tunneling contribution to the forward current transport must be taken into account. Thus, for analyzing the forward characteristic we considered a thermionic field emission (TFE) model [19,33], that is a thermal-assisted tunneling. Fig. 2 (open symbols) reports the semilog plot of the experimental forward J-$V_F$ characteristic of a representative diode acquired at room temperature with the inset depicted a not to scale schematic energy band diagram of a thermal-assisted tunneling transport. In this scheme, $n^+$ and $n^-$ represent the implanted and the epitaxial regions, respectively, while $E_C$ and $E_F$ are the bottom of the conduction band and the Fermi level in the semiconductor. The experimental forward characteristics were fitted by the TFE relationship (continuous line in Fig.2), that is expressed by [33]:

$$J_{TFE} = J_{0,TFE}(V_F) \times \exp\left(q \frac{V_F - J_{TFE}R_{on}}{E_0}\right) \quad (2)$$

where the saturation current $J_{0,TFE}$ ($V_F$) is given by:

$$J_{0,TFE}(V_F) = \frac{A^{**}T}{k_B cosh(qE_{00}/k_BT)} \times$$
$$\sqrt{\pi E_{00}\left(\phi_{B_{TFE}} - \Delta E_F - (V_F - J_{TFE}R_{ON})\right)} \times \exp\left(-\frac{q\Delta E_F}{k_BT} - \frac{\phi_{B_{TFE}} - \Delta E_F}{E_0}\right) \quad (3)$$

and $E_0 = E_{00} \times coth\left(\frac{qE_{00}}{k_BT}\right)$, with q is the elementary charge, $k_B$ is the Boltzmann's constant and T is the absolute temperature. The parameter $E_{00}$ is dependent on the doping concentration $N_D$, according to $E_{00} = \frac{h}{4\pi} \times \sqrt{\frac{N_D}{m^*\varepsilon_{SiC}}}$ with $m^*$ =0.38 $m_0$ the effective mass ($m_0$ is the electron mass) and $\varepsilon_{SiC}$ =9.66 $\varepsilon_0$ the dielectric constant of the semiconductor ($\varepsilon_0$ is the vacuum permittivity) [33,**Errore. Il segnalibro non è definito.**,34,35,36], while $\Delta E_F$ is the difference between the bottom of the conduction band and the semiconductor Fermi level.

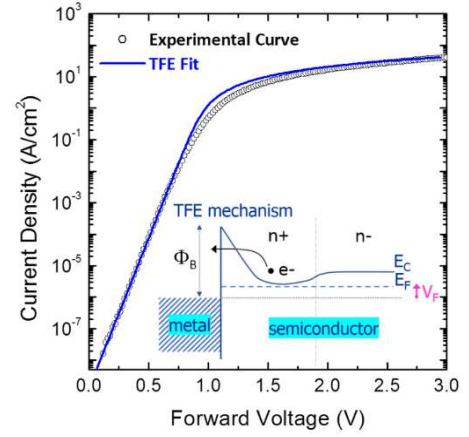

**Figure 2.** Experimental J-$V_F$ curve (open symbols) for Ni/4H-SiC Schottky diode under forward bias at 25 °C and fitting curve according to the TFE model (continuous line). In the inset, schematic energy band diagram for the metal/4H-SiC contact under forward bias, according to the TFE current transport mechanism.

The barrier $\Phi_{B_{TFE}}$ and the doping concentration $N_D$ were determined as parameters of the TFE fit to the experimental J-$V_F$ curve, obtaining $\Phi_{B_{TFE}}$ =1.77 eV and $N_D$=1.97×10$^{19}$ cm$^{-3}$. The ratio $k_BT/qE_{00}$ gives an indication of the relevance of the thermionic emission process with respect to the tunneling one and allows to evaluate which current transport mechanism is predominant for a given doping concentration [34]. Using $N_D$=1.97×10$^{19}$ cm$^{-3}$, the ratio $k_BT/qE_{00}$ is 0.61, which confirms the appropriateness of the TFE model to describe our data.

Furthermore, considering the doping concentration $N_D$=1.97×10$^{19}$ cm$^{-3}$ derived by the TFE fit, a depletion width $W_D$ =10 nm at the Ni/4H-SiC interface can be estimated at zero bias [37]. Such donor concentration is in agreement with the value measured by scanning capacitance microscopy (SCM) of the active dopant profile on a 4H-SiC sample implanted and annealed under the same conditions [23].

Then, the temperature-dependence of the forward and reverse characteristics of the diode was studied to get additional insights into the dominant current transport mechanisms.

Firstly, the temperature-dependence of the forward J-$V_F$ characteristics was monitored between 25 °C and 115 °C (Fig. 3), showing an increase of the current with the temperature. Specifically, by fitting the experimental curves to TFE model for each temperature, we observed a decrease of the barrier $\Phi_{B_{TFE}}$ (from 1.77 to 1.66 eV) with increasing temperature. On the other hand, an almost constant value of doping $N_D$ $(1.96\pm0.02) \times 10^{19}$ cm$^{-3}$ was found as temperature increases. The temperature-dependences of the barrier height $\Phi_{B_{TFE}}$ and the doping concentration $N_D$ are reported as insets in Fig. 3.

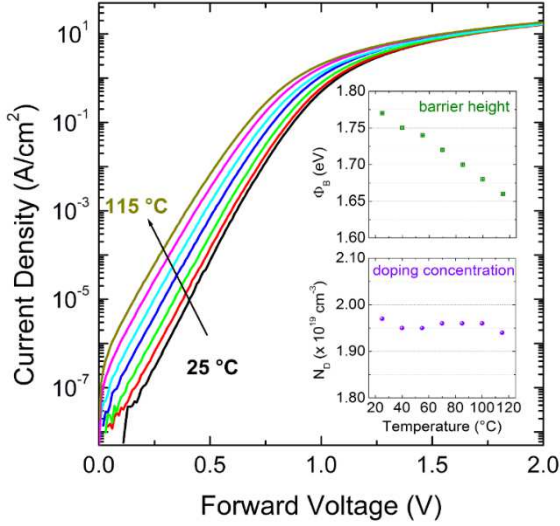

**Figure 3.** Experimental J-V$_F$ curves for Ni/4H-SiC Schottky diode under forward bias at different measurement temperatures in the range 25-115 °C. The insets show the temperature-dependences of the barrier height and doping concentration values, derived by the TFE fits of the forward J-V$_F$ curves at the different measurement temperatures.

Moreover, the effective Richardson's constant determined from the plot ln (Is/T$^2$) vs 1/k$_B$T (where $I_S = AA^*T^2 \times exp\left(-\frac{q\phi_{B_{TE}}}{k_BT}\right)$ is the saturation current of eq. 1), was 4.7×10-4 A·cm−2·K-2) ( not shown here). This experimental value is lower than the expected one of 146 A·cm$^{-2}$·K$^{-2}$ for 4H-SiC [30]. This discrepancy has been often attributed to deviations from the thermionic emission model and/or to lateral inhomogeneity of the barrier [30,38].

Fig. 4a reports the reverse current density-voltage characteristic (J-V$_R$) of the Ni/4H-SiC Schottky diode acquired at various temperatures. As can be seen, the reverse current density increases from tenths of nA/cm$^2$ up to some units of A/cm$^2$ with increasing reverse bias. Noticeably, while the reverse current exhibits a clear dependence on the measurement temperature for lower voltage, it becomes almost independent of the temperature at higher bias values.

The field emission (FE) regime through a metal/semiconductor interface predicts only a weak temperature dependence of the reverse current, with the reverse current density expressed as [33]

$$J_{FE}(V_R) = \frac{qA^{**}T^2\pi E_{00} \times exp\left(\frac{2\phi_b^{3/2}}{3E_{00}(\phi_B - V_R)^{1/2}}\right)}{k_BT[\phi_B/(\phi_B - V_R)]^{1/2}sin\{\pi k_BT[\phi_B/(\phi_B - V_R)]^{1/2}/E_{00}\}} \quad (4)$$

with the meaning of symbols as mentioned above.

At room temperature (inset of Fig. 4a), the FE model could fit the experimental data with a barrier height $\Phi_B$=1.77e V and a doping concentration N$_D$=1.0×10$^{19}$cm$^{-3}$. In this case, an effect of the series resistance at high current level (high reverse voltage) cannot not be ruled out. Moreover, the FE model could not well describe the experimental curves at all measurement temperatures. Hence, it is possible to argue that an additional mechanism must be taken into account to explain the more pronounced temperature behaviour of the reverse current at low voltage.

Fig. 4b reports an Arrhenius' plot of the current density for three representative reverse biases (at 1V, 3V and 8V). From these plots, it was possible to determine the activation energy that varies from of 0.387eV at 1V down to 0.205eV at 3V. Interestingly, the Arrhenius' plot gives a much lower activation energy of 0.052eV at 8V, where the current is almost independent of temperature. This latter is consistent with the prevalence of a tunnelling mechanism coexisting with the series resistance contribution of the epilayer.

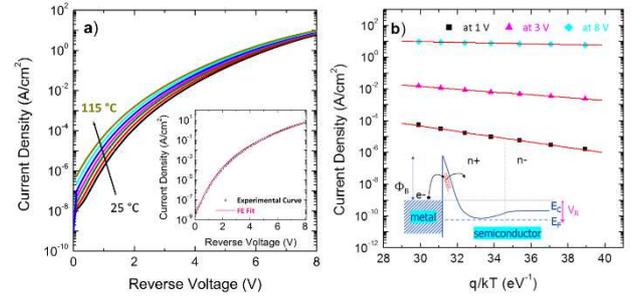

**Figure 4.** (a) Experimental J-V$_R$ characteristics of the Ni/4H-SC contact under reverse bias in the temperature range 25-115 °C. The inset reports the reverse characteristic acquired at room temperature (open symbols) and the fitting curve according to the FE regime (continuous line). (b) Arrhenius' plot of the current density at three different biases (1V, 3V and 8V), giving activation energy of 0.387eV at 1V and 0.205eV at 3V; the current is almost independent of the temperature at 8V. In the inset, schematic energy band diagram for trap-assisted tunneling contribution to the current transport under moderate reverse bias.

As observed in literature [19,39,40,41,42,43,44], a large variety of defects can be induced in 4H-SiC by ion-implantation and post thermal treatments, introducing energy levels within the band gap of the material, which can have an impact on the leakage current of Schottky diodes. Plausibly, the defects in our high dose P-ion implanted sample induce energy levels that can assist the current transport in the moderate reverse bias range, explaining the temperature-dependence of the reverse characteristics (inset Fig.4b). At higher voltages, the direct tunneling FE becomes progressively dominant, due to the barrier thickness reduction.

A possible application of these results in a real device, was considered by performing a numerical simulation of the potential distribution in a JBS diode structure. The JBS consists in embedding p+-type regions (usually achieved by ion implantation) within an n-type Schottky area. In this layout, the leakage current of the Schottky contact on a n+-doped surface region can be mitigated by the lateral depletion of the p+-n junctions. The numerical simulation was carried out using a drift diffusion solver implemented in the open source FEniCS platform [45] and employing a heavily doped n+-type region below the Schottky metal, is shown in Fig. 5 for a reverse bias of -10V. The JBS structure (inset of Fig. 5) assumes ideally 0.6 $\mu$m-thick rectangular p+-regions ($N_A = 2\times 10^{19}$ cm$^{-3}$) placed around a 1.4 $\mu$m wide Schottky contact formed on a thin 50 nm n+-region ($N_D = 1\times 10^{19}$ cm$^{-3}$) over a n- drift layer ($N_D = 1\times 10^{16}$ cm$^{-3}$). The experimental value of the Schottky barrier height is used on the n+-region, while ohmic contacts are assumed in the p+- regions.

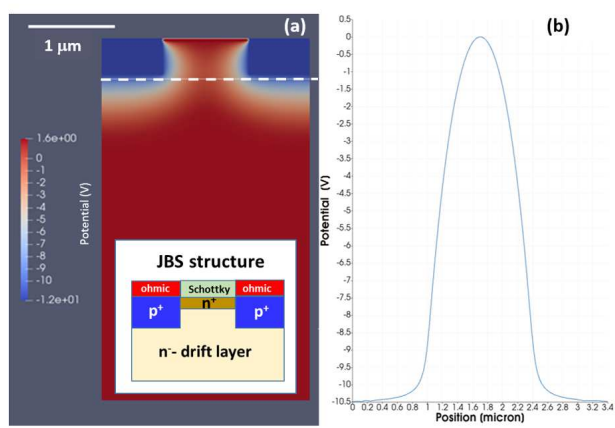

**Figure 5.** (a) Numerical study of the potential distribution for a JBS structure (schematically depicted in the inset) employing a heavily doped n+-type region below the Schottky metal, for a reverse bias of -10V. The potential reference in the bulk n- epi region is aligned to its quasi-Fermi energy. (b) Cut line of the potential distribution 50nm above the p+-regions, showing the pinch-off effect.

Evidently, a full depletion (pinch-off) of the region below the Schottky metal is generated by the p+/n- lateral junction under this reverse bias condition, which should suppress the leakage current of the device. Obviously, this action is gradual with increasing the reverse bias, with an overall benefit on the diode features. On the other hand, the lower turn-on voltage under forward bias guarantees a low power consumption compared to the case of a conventional JBS structure without the n+-type region below the Schottky contact.

It is worth noting that in such conventional devices, the pinch-off condition can be achieved already at a lower reverse bias ($\approx$ -6V) However, with an appropriate device layout (i.e., width of the Schottky and Ohmic regions), the proposed method can be promising to control the barrier in JBS diodes.

## 4. Conclusions

In conclusion, the electrical behavior of a Ni Schottky barrier formed onto heavily ($N_D > 10^{19}$ cm$^{-3}$) doped n-type implanted 4H-SiC was investigated, elucidating the current transport mechanisms in both forward and reverse bias. An accurate analysis of the current transport showed that current injection in forward bias is described by a thermionic-field emission mechanism. Under reverse bias, the temperature behavior of the leakage current is affected by the presence of implantation-induced damage at low bias, while a pure field emission mechanism is approached with increasing the bias. A numerical study of the potential distribution in a JBS diode demonstrated the possibility to apply this process to improve the performance of real 4H-SiC Schottky

## Acknowledgements

The authors would like to acknowledge STMicroelectronics for providing 4H-SiC implanted samples and S. Di Franco (CNR-IMM) for technical assistance in the diodes fabrication. Part of this research activity has been carried out in the framework of the European ECSEL JU project REACTION (Grant Agreement No. 783158), using the facilities of the Italian Infrastructure Beyond-Nano.